\documentclass[aps,pra,eqsecnum,twocolumn]{revtex4}
\usepackage{graphicx}
\usepackage{bm}
\usepackage{amsmath,amssymb,amsthm,dsfont,bm,enumitem}
\usepackage{color}
\usepackage{wasysym}
\usepackage{ulem}
\usepackage[dvipsnames]{xcolor}

\usepackage[applemac]{inputenc}

\usepackage[colorlinks=true,urlcolor=blue,citecolor=blue,linkcolor=blue]{hyperref}

\begin{document}
\preprint{}
\title{Generalized measurements for Bell tests in different probability spaces}
\author{Alfredo Luis}
\email{alluis@fis.ucm.es}
\homepage{https://sites.google.com/ucm.es/alfredo/inicio}
\affiliation{Departamento de \'{O}ptica, Facultad de Ciencias
F\'{\i}sicas, Universidad Complutense, 28040 Madrid, Spain}
\date{\today}

\begin{abstract}
Bell tests are of profound statistical nature. Besides physical considerations, the proper understanding of their implications should involve detailed statistical analyses. In this regard, recent works have shown that their consequences and interpretations depend on the probability space adopted. Some other recent works have also shown that generalized measurements may allow to further exploit the statistics of Bell-like tests. Following these ideas, in this work we show that one and the same experimental arrangement can provide a practical scheme valid for two very different probability spaces. Moreover, we show that this allows the introduction of novel Bell tests that are not possible in more standard approaches.
\end{abstract}
\maketitle

\section{Introduction}

Bell tests provide a powerful tool for investigating fundamental concepts at the very heart of the quantum theory \cite{JB64,LB90,WW93,AF82,AR15,BKO16,CH74,CHSH69,MC88,AK00,HP04,AM08}. Deep down, Bell tests are a subject of statistical analysis \cite{SGH21,SH22,AA24,MA84,TN11,JCh17}. In this regard, recent works have shown that the their consequences and interpretations may heavily depend on the probability space adopted \cite{SGH21,SH22}.

However, the statistical analysis of Bell tests cannot be carried out in the most standard approaches. This is because a key feature of such tests is that they unavoidable involve the measurement of incompatible observables at each party. This requires then to combine results from different incompatible experimental realizations, having different physical contexts and performed at different times. This prevents their joint statistical treatment beyond mean values. 

At this point, recent  works have demonstrated the importance of generalized measurements to address the statistics of Bell tests \cite{MAL20,VRA23,GP25,AL25a,AL25b}. These are practical schemes where all the observables are measured at once in a single experimental arrangement \cite{MAL22,MM89,WMM02,WMM14,PB87,AL16,LM17,YLLO10,AL25c}.

This offers a plenty of possibilities regarding practical and interpretational issues \cite{GP25}. For example, this includes applying Bell tests after single measurements \cite{AL25a,AHQ20}, as well as  addressing the probability that a given state may violate the Bell bounds \cite{VRA23,GP25,AL25b}. 

Another advantage of generalized measurements is presented in this work. This is that one and the same experimental arrangement can provide a practical scheme valid for the two main probability spaces analyzed in Refs. \cite{SGH21,SH22}. This can help to exploit all the statistical possibilities of Bell tests. In particular we show that this allows the introduction of some new Bell tests that are not possible withing more standard approaches. 

\section{A generalized measurement}

We address a quantum-optical realization of Bell tests, where the two subsystems A and B are  two-mode single photons. Each photon is described by a two-dimensional Hilbert space expressing transversal polarization. For example for subsystem A the most general on-photon pure state is of the form
\begin{equation}
\label{input}
    |\psi \rangle = \mu |1 \rangle_x |0 \rangle_y +  \nu |0 \rangle_x |1 \rangle_y \equiv  \begin{pmatrix} \mu \\\nu \end{pmatrix} ,
\end{equation}
where $|n\rangle_j$ are photon-number states in the linearly polarized field modes $j=x,y$ with complex-amplitude operators $a_j$, and $\mu,\nu$ are arbitrary complex number properly normalized $|\mu|^2+|\nu|^2=1$. 

\begin{figure}
\centering
\includegraphics[width=8cm]{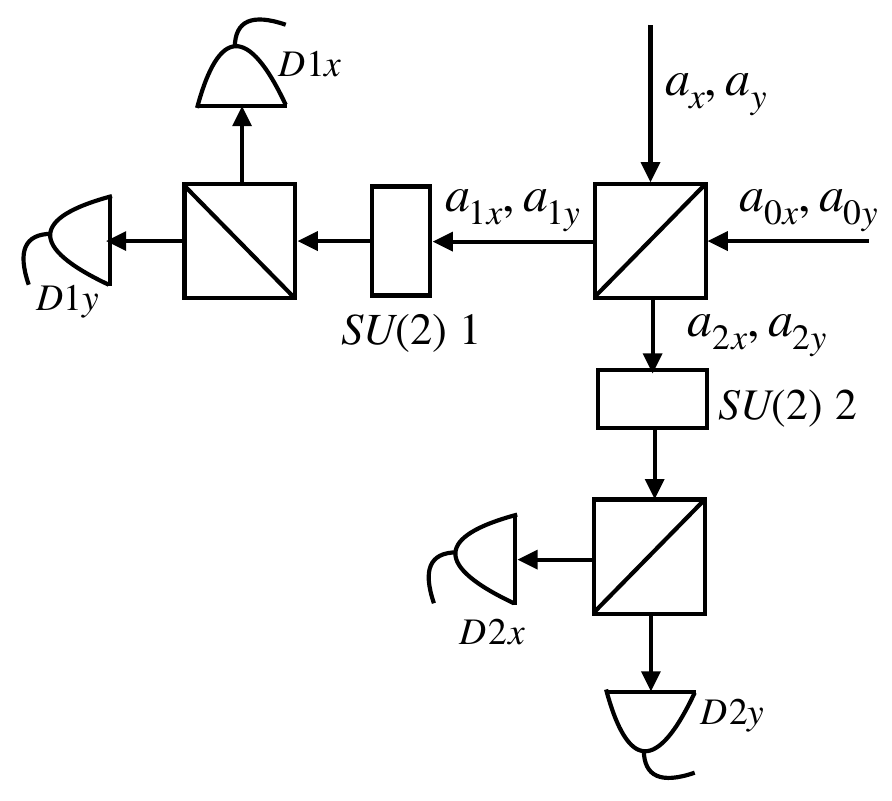}
\label{experimento}
\caption{Sketch of the generalized measurement showing the arrangement in one of the subsystems, say A.}
\end{figure}

Let us describe the generalized measurement as sketched in Fig. 1. The input signal modes are described by complex-amplitude operators $a_x, a_y$, are mixed at a beam splitter with other two inputs modes, with complex-amplitude operators $a_{0x}, a_{0y}$, which will be always in the vacuum state. These auxiliary modes  provide the extra room allowing the joint simultaneous measurement of different observables. The beam splitter is lossless and conveniently described by real transmission and reflection coefficients $t_x, t_y, r_x, r_y$, in principle different for the $x$ and $y$ polarizations, always with $t_x^2+r_x^2 = t_y^2+r_y^2 =1$. With this, the complex-amplitude operators of the modes leaving the beam splitter are related to the input ones as
\begin{equation}
\label{mt1}
    a_{1x} = t_x a_{0x} + r_x a_x, \quad  a_{1y} = t_y a_{0y} + r_y a_y, 
\end{equation}
for the output labeled as 1, and similarly for the output labeled as 2
\begin{equation}
\label{mt2}
    a_{2x} = t_x a_x - r_x a_{0x}, \quad a_{2y} = t_y a_y - r_y a_{0y} .
\end{equation}
The field state in these output modes after Eqs. (\ref{input}), (\ref{mt1}), and (\ref{mt2}) is then
\begin{eqnarray}
\label{output}
  &  |\psi \rangle = \mu r_x |1 \rangle_{1x} |0 \rangle_{1y} |0 \rangle_{2x}|0 \rangle_{2y} & \nonumber \\
   & & \nonumber \\
 & + \mu t_x  |0 \rangle_{1x} |0 \rangle_{1y} |1 \rangle_{2x}|0 \rangle_{2y}  & \nonumber \\
  & & \nonumber \\
  & + \nu r_y |0\rangle_{1x} | 1\rangle_{1y} |0 \rangle_{2x}|0 \rangle_{2y} & \nonumber \\
   & & \nonumber \\
 & + \nu t_y |0 \rangle_{1x} |0 \rangle_{1y} |0 \rangle_{2x}|1 \rangle_{2y}   .& 
\end{eqnarray}

After the beam splitter the output modes experience a polarization transformations represented by the SU(2) boxes in Fig. 1. They are then followed by polarizing beam splitters separating the $x$ and $y$ modes, directing them to photon detectors $D1x, D1y$ and $D2x, D2y$. Given the one-photon nature of each subsystem, in each run of the experiment only one of the of the four detectors $D1x, D1y,  D2x, D2y$ clicks. The click of detectors $D1x$ and $D1y$  occurs with probabilities given by projection of the field state (\ref{output}) on the states
\begin{equation}
     |\varphi_{1x} \rangle =|\Omega_1,1 \rangle |0 \rangle_{2x}|0 \rangle_{2y} , \quad
     |\varphi_{1y} \rangle = |\Omega_1,-1 \rangle|0 \rangle_{2x}|0 \rangle_{2y},
\end{equation}
where
\begin{equation}
 |\Omega_1,1 \rangle= \cos \frac{\theta_1}{2}  |1 \rangle_{1x} |0 \rangle_{1y} +  e^{i \phi_1} \sin \frac{\theta_1}{2} |0 \rangle_{1x} |1 \rangle_{1y} ,
\end{equation}
and
\begin{equation}
 |\Omega_1,-1 \rangle = -\sin \frac{\theta_1}{2}  |1 \rangle_{1x} |0 \rangle_{1y} +  e^{i \phi_1} \cos \frac{\theta_1}{2} |0 \rangle_{1x} |1 \rangle_{1y} ,
 \end{equation}
where $\Omega_1 = (\theta_1,\phi_1)$ are polarization parameters associated to the polarization transformation SU(2) 1 in Fig. 1. Similarly regarding the clicks in modes $a_2$, that is 
\begin{equation}
|\varphi_{2x} \rangle_ = |0 \rangle_{1x}|0 \rangle_{1y} |\Omega_2, 1 \rangle , \quad
|\varphi_{2y} \rangle = |0 \rangle_{1x}|0 \rangle_{1y} |\Omega_2, -1 \rangle ,
\end{equation}
where 
\begin{equation}
|\Omega_2, 1 \rangle = \cos \frac{\theta_2}{2}  |1 \rangle_{2x} |0 \rangle_{2y} +  e^{i \phi_2} \sin \frac{\theta_2}{2} |0 \rangle_{2x} |1 \rangle_{2y} ,
\end{equation}
and 
\begin{equation}
|\Omega_2, -1 \rangle = -\sin \frac{\theta_2}{2}  |1 \rangle_{2x} |0 \rangle_{2y} +  e^{i \phi_2} \cos \frac{\theta_2}{2} |0 \rangle_{2x} |1 \rangle_{2y},
\end{equation}
where $\Omega_2 = (\theta_2,\phi_2)$ are polarization parameters associated to the polarization transformation SU(2) 2 in Fig. 1.

With all this, the statistics of these clicks is given by projection of the input state (\ref{input}) in modes $a_{x,y}$ in the unnormalized states 
\begin{equation}
\label{ds1}
    |\psi_{1x} \rangle  \equiv  \begin{pmatrix} r_x \cos \frac{\theta_1}{2} \\  r_y e^{i \phi_1} \sin \frac{\theta_1}{2}  \end{pmatrix} , \quad
|\psi_{1y} \rangle  \equiv  \begin{pmatrix} - r_x \sin \frac{\theta_1}{2} \\ r_y e^{i \phi_1} \cos \frac{\theta_1}{2}  \end{pmatrix} ,
\end{equation}
and
\begin{equation}
\label{ds2}
    |\psi_{2x} \rangle  \equiv  \begin{pmatrix} t_x \cos \frac{\theta_2}{2} \\ t_y e^{i \phi_2} \sin \frac{\theta_2}{2}  \end{pmatrix} , \quad
|\psi_{2y} \rangle  \equiv  \begin{pmatrix} - t_x \sin \frac{\theta_2}{2} \\ t_y e^{i \phi_2} \cos \frac{\theta_2}{2}  \end{pmatrix} .
\end{equation}
{\it Mutatis mutandis}, this can be translated to the clicks of the four detectors at subsystem B for some other polarization transformations and parameters  $\Omega^\prime_1 = (\theta^\prime_1,\phi^\prime_1)$ and $\Omega^\prime_2 = (\theta^\prime_2,\phi^\prime_2)$ 

With this we can construct a great deal of observables as suitable linear combinations of these four nonorthonormal projectors, whose statistics will be linear combination of these clicks probabilities. Here we will consider just two examples related to two different cases of probability spaces as introduced in Refs. \cite{SGH21,SH22}.

\section{Probability space 1}

Following the approach in Refs. \cite{SGH21,SH22} we can fully describe each subsystem with two dichotomic variables labelling the four possible outcomes, say $j,\alpha = \pm 1$ for subsystem A, and equivalently $k ,\beta = \pm 1$ for subsystem B. The variables $\alpha,\beta$ design the observable we are observing at each subsystem, two observables $A_\alpha$ in A and other two $B_\beta$ in B, while $j, k$ represent the outcomes. This can be conveniently implemented within our general scheme by using a nonpolarizing beam splitter $t_x=t_y=t_A$ and  $r_x=r_y=r_A$, followed by a suitably adscription of meaning to the detectors clicks. Let us say that the clicks at detectors $D1x, D1y$ constitute the two outputs of an observable $A_1$. Likewise, the clicks at detectors $D2x, D2y$ constitute the two outputs of another observable $A_{-1}$. That is 
\begin{eqnarray}
& \text{click of detector } D1x \rightarrow  \alpha=1, j=1 ,& \nonumber \\ & & \nonumber \\
& \text{click of detector } D1y \rightarrow  \alpha=1, j=-1 ,& \nonumber \\ & & \nonumber \\ 
& \text{click of detector } D2x \rightarrow  \alpha=-1, j=1 ,& \nonumber \\ & & \nonumber \\ 
& \text{click of detector } D2y \rightarrow  \alpha=-1, j=-1 .& \nonumber  
\end{eqnarray}
In this case, the projecting states (\ref{ds1}) determining the statistics by projection  on the input state (\ref{input}) are
\begin{equation}
\label{}
    |\psi_{1x} \rangle  =  r_A \begin{pmatrix}\cos \frac{\theta_1}{2} \\  e^{i \phi_1} \sin \frac{\theta_1}{2}  \end{pmatrix} , \quad
|\psi_{1y} \rangle  =  r_A \begin{pmatrix} - \sin \frac{\theta_1}{2} \\  e^{i \phi_1} \cos \frac{\theta_1}{2}  \end{pmatrix} ,
\end{equation}
which are orthogonal eigenvectors of the $A_1$ operator 
\begin{equation}
    A_1 = \vec{S}_1 \cdot {\vec{\sigma}} ,
\end{equation}
where $\vec{S}_1$ is the following three-dimensional real vector, actually a point of the Poincar\'e sphere, 
\begin{equation}
    \vec{S}_1 = (\sin \theta_1 \cos \phi_1, \sin \theta_1 \sin \phi_1, \cos \theta_1 ) ,
\end{equation}
and $\vec{\sigma}$ a three-dimensional vector with the Pauli matrices 
\begin{equation}
    \vec{\sigma} = (\sigma_1, \sigma_2, \sigma_3) .
\end{equation}
Similarly, the projecting states (\ref{ds2}) are
\begin{equation}
    |\psi_{2x} \rangle  = t_A \begin{pmatrix}  \cos \frac{\theta_2}{2} \\  e^{i \phi_2} \sin \frac{\theta_2}{2}  \end{pmatrix} , \quad
|\psi_{2y} \rangle  =  t_A \begin{pmatrix} - \sin \frac{\theta_2}{2} \\  e^{i \phi_2} \cos \frac{\theta_2}{2}  \end{pmatrix} ,
\end{equation} 
which are the orthogonal eigenstates of the observable $A_{-1}$ 
\begin{equation}
    A_{-1} = \vec{S}_{-1} \cdot {\vec{\sigma}} ,
\end{equation}
with 
\begin{equation}
    \vec{S}_{-1} = (\sin \theta_2 \cos \phi_2, \sin \theta_2 \sin \phi_2, \cos \theta_2 ) .
\end{equation}

The joint probability for these variables is given by
\begin{equation}
    p(j,\alpha) = \mathrm{tr} \left [ \rho_A \Delta_A (j, \alpha) \right ]
\end{equation}
where $\rho_A$ is the subsystem A density matrix, and the corresponding positive operator-values measure (POVM) is
\begin{equation}
\label{Dja}
\Delta_A (j, \alpha) = p(\alpha) \frac{1}{2} \left ( \sigma_0 + j \vec{S}_\alpha \cdot \vec{\sigma} \right ),
\end{equation}
where $p(\alpha)$ is 
\begin{equation}
\label{pa}
    p(\alpha) = \frac{1+\alpha }{2} r_A^2 + \frac{1 - \alpha}{2} t_A^2 ,
\end{equation}
and $\sigma_0$ is the $2 \times 2$ identity matrix.

Equivalently for subsystem B regarding observables $B_\beta$ at the corresponding outputs of the beam splitter mixing with vacuum, that is 
\begin{equation}
\Delta_B (k, \beta) = p(\beta) \frac{1}{2} \left ( \sigma_0 + k \vec{S}_\beta \cdot \vec{\sigma} \right ),
\end{equation}
where $p(\beta)$ is 
\begin{equation}
\label{pb}
    p(\beta) = \frac{1+\beta }{2} r_B^2 + \frac{1 - \beta}{2} t_B^2 ,
\end{equation}
and here matrices $\vec{\sigma}$ operate in the corresponding B space.

\subsection{Marginals}

It is worth having a look on the marginal distribution and probabilities for the $j,\alpha$ variables. For variable $\alpha$ we have a trivial marginal POVM, this is 
\begin{equation}
\label{Da}
    \Delta (\alpha) = \sum_{j=\pm 1}  \Delta_A (j,\alpha)=p(\alpha) \sigma_0 ,
\end{equation}
where we can see that the $p(\alpha)$ introduced in Eq. (\ref{pa}) is actually the probability of $\alpha$, which turns out to be independent of the field state. Regarding the marginal for the $j$ variable we have  
\begin{equation}
\label{Dj}
    \Delta (j) = \sum_{\alpha=\pm 1} \Delta_A (j, \alpha) = \frac{1}{2} \left ( \sigma_0 + j \vec{S}_A \cdot  \vec{\sigma} \right ),
\end{equation}    
where 
\begin{equation}
\label{SA}
\vec{S}_A = \sum_{\alpha=\pm 1} p(\alpha) \vec{S}_\alpha = r_A^2 \vec{S}_1 + t_A^2 \vec{S}_{-1} .
\end{equation}

\subsection{Conditional probabilities, POVMs, and states}

With the joint distribution $p(j,\alpha)$ and its marginals $p(\alpha)$, $p(j)$, we may construct conditional probabilities such that, for example,  
\begin{equation}
    p(j|\alpha ) = \frac{ p(j,\alpha )}{ p(\alpha )} .
\end{equation}
Let us note that we can express $\Delta_A (j, \alpha)$ as
\begin{equation}
    \Delta_A (j, \alpha) = p(\alpha) \Delta (j|\alpha), 
\end{equation}
where 
\begin{equation}
   \Delta (j |\alpha) =  \frac{1}{2} \left ( \sigma_0 + j \vec{S}_\alpha \cdot \vec{\sigma} \right ) ,
\end{equation}
are the projection-valued measure corresponding to the exact measurement of the observables $A_\alpha$. Then
\begin{equation}
    p(j|\alpha ) = \mathrm{tr} \left [ \rho_A \Delta (j |\alpha) \right ] ,
\end{equation}
is the exact statistics of the observable $A_\alpha$ in the state $\rho_A$. It is worth noting that, strictly speaking, we planned this detection scheme as a noisy joint determination of incompatible observables. But it turns out that in this case the additional noise is so simple that it is completely removed by a proper and simple derivation of conditional probabilities. Deep down this scheme is equivalent to a fully random choice of the observable to be measured in each subsystem with probabilities $p(\alpha,\beta)$. Moreover we can note that after Eqs. (\ref{pa}) and (\ref{Da}), we get $p(\alpha,\beta) = p(\alpha) p (\beta)$ so Bell-type correlations are not due to observable-choice correlations, since they are fully independent. 

Moreover, following the work in Ref. \cite{PL22} we may ask whether the Gleason theorem applies to conditional probabilities. This is that whether there are legitimate density matrices $\rho_\alpha$ such that
\begin{equation}
    p(j|\alpha ) = \mathrm{tr} \left [ \rho_\alpha \Delta (j ) \right ] .
\end{equation}
Expressing $\rho_\alpha$ and the reduced density matrix in subsystem A $\rho_A$ and in terms of Pauli matrices 
\begin{equation}
\label{rho}
\rho_A =  \frac{1}{2} \left ( \sigma_0 + \vec{s}_A \cdot  \vec{\sigma} \right ), \quad
    \rho_\alpha =  \frac{1}{2} \left ( \sigma_0 + \vec{s}_\alpha \cdot  \vec{\sigma} \right ), 
\end{equation}
we have that the only condition on $\vec{s}_\alpha$ is that its projection on $\vec{S}_A$ must satisfy the relation 
\begin{equation}
    \vec{s}_\alpha \cdot \vec{S}_A = \vec{s}_A \cdot \vec{S}_\alpha .
\end{equation}
This condition cannot be always satisfied since for example $\vec{S}_A$ can have a vanishing modulus depending on the choice of $\vec{S}_{\pm 1}$ and $t_A$, while $\vec{s}_\alpha$ have at  most unit modulus $|\vec{s}_\alpha | \leq 1$. More, specifically the necessary and sufficient condition for the existence of $\rho_\alpha$ is that $|\vec{s}_A \cdot \vec{S}_\alpha |\leq |\vec{S}_A|$.

\bigskip

Likewise we may as well consider the conditional probabilities 
\begin{equation}
\label{pcaj}
    p(\alpha|j) = \frac{ p(j,\alpha )}{ p(j )} .
\end{equation}
In this case, there is no $\Delta (\alpha|j)$ such that 
\begin{equation}
\label{npP}
    p(\alpha|j ) = \mathrm{tr} \left [ \rho_A \Delta (\alpha |j) \right ] .
\end{equation}
This is clear after using again the parametrization $\vec{s}_A$ for the subsystem A density matrix in Eq. (\ref{rho}) and properly using Eqs. (\ref{Dja}) and (\ref{Dj}) we get 
\begin{equation}
    p(\alpha|j) = p(\alpha) \frac{1+j \vec{S}_\alpha \cdot \vec{s}_A }{1+ j \vec{S}_A\cdot \vec{s}_A  } ,
\end{equation}
so Eq. (\ref{npP}) cannot take place. 

Here again we may ask whether the Gleason theorem works if there are legitimate density matrices $\rho_j$ such that
\begin{equation}
    p(\alpha |j) = \mathrm{tr} \left [ \rho_j \Delta (\alpha ) \right ] .
\end{equation}
But clearly there is no $\rho_j$ able to satisfy this relation because $\Delta (\alpha )$ is trivial and for all $\rho_j$ 
\begin{equation}
\label{pcaj=}
   \mathrm{tr} \left [ \rho_j \Delta (\alpha ) \right ] =p (\alpha) ,
\end{equation}
but in general $ p(\alpha |j) \neq p (\alpha)$. 

\subsection{Joint probabilities}

Finally, the joint probabilities for the whole set of variables is 
\begin{equation}
    p(j,k,\alpha,\beta ) = \mathrm{tr} \left [ \rho \Delta_A (j, \alpha) \otimes \Delta_B (k, \beta) \right ], 
\end{equation}
where $\rho$ is the density matrix for the complete system.

\subsection{Standard Bell test}

With the above ingredients, suitable Bell tests can be constructed. For example we can begin with  the Bell test that emerge in this framework from the proper combination of conditional probabilities
\begin{equation}
p(j,k|\alpha,\beta) = \frac{p(j,k, \alpha,\beta)}{p(\alpha,\beta)} .
\end{equation}
for a fixed, given outcome $j,k$ by gathering the two possibilities for $\alpha$ and $\beta$ in the usual way
\begin{eqnarray}
\label{CBt}
 &   C = p(j,k|\alpha,\beta) - p(j,k|\alpha,-\beta) + p(j,k|-\alpha,\beta) &\nonumber \\
    & & \nonumber \\ 
    &+ p(j,k|-\alpha,-\beta) - p(j|-\alpha) - p(k|\beta) . &
\end{eqnarray}
As shown in Ref. \cite{SGH21,SH22} regarding a hidden variables $\lambda$ model assuming locality 
\begin{equation}
\label{locality}
    p(j,k|\alpha,\beta,\lambda ) =  p(j|\alpha,\lambda ) p(k|\beta,\lambda ),  
\end{equation}
the following bound can be derived
\begin{equation}
\label{bBt}
    0 \ge C \geq -1 .
\end{equation}
It is worth noting that all these conditional probabilities are exact joint statistics of the corresponding $\alpha,\beta$ observables measured all them in a single arrangement. Because of this, in the derivation of Eq. (\ref{bBt}) there is no need to impose a setting independence condition of the form, say, $p(\lambda |\alpha,\beta ) = p(\lambda)$ because there is a single context for the complete test. 

\subsection{Dual and mixed Bell test}

One of the virtues of this probability-space formalism is that it clearly states that the ingredients of the Bell test (\ref{bBt}) are not probabilities, but conditional probabilities  \cite{SGH21,SH22,AK14}. And as we have shown above, the conditional probabilities that can be derived from the joint statistics does not reduce to the standard formulation. So let us derive some interesting alternatives.

We may say that the $j, k$ variables should be on an equal footing with  $\alpha,\beta$, variables. Then it may be worth inquiring whether some new Bell tests may be derived by exchanging the roles of the $j,k$ and $\alpha, \beta$ variables within this very same probability space. That is, we consider now the conditional probabilities 
\begin{equation}
p(\alpha,\beta |j,k) = \frac{p(j,k, \alpha,\beta)}{p(j,k)} ,
\end{equation}
and in general $p(j,k) \neq p(j)p(k)$. In this case after assuming some equivalent to the locality condition in Eq. (\ref{locality}) 
\begin{equation}
    p(\alpha,\beta|j,k,\lambda ) =  p(\alpha | j,\lambda ) p(\beta |k,\lambda ),  
\end{equation}
we would have the analogous Bell-type bound 
\begin{equation}
    0 \ge C^\prime \geq -1 .
\end{equation}
where
\begin{eqnarray}
&  C^\prime = p(\alpha,\beta|j,k) - p(\alpha,\beta|j,-k ) + p(\alpha,\beta|-j ,k) & \nonumber \\
& & \nonumber \\ &+ p(\alpha,\beta |-j, -k ) - p(\alpha|-j ) - p(\beta|k) . & 
\end{eqnarray}

\bigskip

We may as well consider mixed conditional probabilities, say 
\begin{equation}
p(j,\beta |\alpha,k) = \frac{p(j,k, \alpha,\beta)}{p(k,\alpha)} ,
\end{equation}
and in this case $p(k,\alpha)=p(k)p(\alpha)$. After assuming some equivalent to the locality condition in Eq. (\ref{locality}) 
\begin{equation}
    p(j,\beta|\alpha,k,\lambda ) =  p(j | \alpha,\lambda ) p(\beta |k,\lambda ),  
\end{equation}
we would have the analogous Bell-type bound 
\begin{equation}
    0 \ge C^{\prime \prime} \geq -1 .
\end{equation}
where
  \begin{eqnarray}
&  C^{\prime \prime} = p(j,\beta|\alpha,k) - p(j,\beta|\alpha,-k) + p(j,\beta|-\alpha,k) & \nonumber \\
& & \nonumber \\ &+p(j,\beta|-\alpha, -k) - p(j|-\alpha ) - p(\beta|k) . & 
\end{eqnarray},

\bigskip

\section{Probability space 2}

Next we consider the practical implementation of a different probability space, the probability space 2 in Refs. \cite{SGH21,SH22},  using the very same measuring scheme in Fig. 1, but via a different rearrangement of outcomes, as a generalized version of the eight-port homodyne detector already considered in Ref. \cite{LM17}. Let us define the following two dichotomic variables $j,k = \pm 1$ as follows, for subsystem A
\begin{eqnarray}
& j=1, k=1 \rightarrow \text{click of detector } D1x ,& \nonumber \\ & & \nonumber \\ & j=-1, k=-1 \rightarrow \text{click of detector } D1y ,& \nonumber \\ & &  \\ & j=1, k=-1 \rightarrow \text{click of detector } D2x ,& \nonumber \\ & & \nonumber \\ & j=-1, k=1 \rightarrow \text{click of detector } D2y .& \nonumber \\          
\end{eqnarray}
This is to say we construct a POVM $\Delta_A (j,k)$ in terms of projections on the unnormlized, nonorthogonal detector states (\ref{ds1}), (\ref{ds2}) as follows 
\begin{equation}
\Delta_A (1,1) = |\psi_{1x} \rangle \langle \psi_{1x} |, \quad \Delta_A (-1,-1) = |\psi_{1y} \rangle \langle \psi_{1y} |  ,
\end{equation}
and 
\begin{equation}
\Delta_A (1,-1) = |\psi_{2x} \rangle \langle \psi_{2x} |, \quad \Delta_A (-1,1) = |\psi_{2y} \rangle \langle \psi_{2y} |  .
\end{equation}

\subsection{Particular case}

General expressions for this POVM $\Delta_A (j,k)$ are rather involved and not very indicative, so for the sake of illustration let us consider the particular  simple but meaningful enough case with $t_x= r_y = t_A$, $t_y= r_x = r_A$, $\theta_1=\theta_2 = \pi/2$, $\phi_1 = - \phi_2 = \pi/4$ to get 
\begin{eqnarray}
& \Delta_A (j,k) = \frac{1}{4} \left ( \sigma_0 + j  \gamma_X \vec{S}_X \cdot \vec{\sigma} + k \gamma_Y \vec{S}_Y \cdot \vec{\sigma}  \right .& \nonumber \\
 & & \nonumber \\
 & \left . +jk \gamma_{XY} \vec{S}_{XY} \cdot \vec{\sigma}  \right ) , & 
\end{eqnarray}
where 
\begin{equation}
    \vec{S}_X =   \left ( 1, 0, 0 \right) , \quad \gamma_X = \sqrt{2} rt ,
\end{equation}
\begin{equation}
\vec{S}_Y = \left ( 0,1, 0 \right) , \quad \gamma_Y = \sqrt{2} rt ,
\end{equation} 
\begin{equation}
\vec{S}_{XY} =\left ( 0,0,1 \right) ,  \quad \gamma_{XY} = r^2- t^2.
\end{equation}
It is worth noting that the $\gamma$ factors satisfy the following link
\begin{equation}
    \gamma_X^2+\gamma_Y^2 +\gamma_{XY}^2 = 1. 
\end{equation}
In particular we may give the same importance to the three observables $X$, $Y$ and $Z$, by choosing $r,t$ such that $\gamma_X = \gamma_Y =\gamma_{XY} = 1/\sqrt{3}$ in which case $\Delta_A (j,k)$ becomes the POVM for minimal qubit tomography studied in Ref. \cite{REK04}. 

\bigskip

\subsection{Marginals and noisy measurement}

The  marginals determine the observables $X$, $Y$ corresponding to variables $j,k$. Say, in terms of POVMs
\begin{equation}
\Delta_X (j) = \sum_{k=\pm 1} \Delta_A (j,k) = \frac{1}{2} \left ( \sigma_0 + j  \gamma_X \vec{S}_X \cdot \vec{\sigma}  \right ) , 
\end{equation}
and
\begin{equation}
\Delta_Y (k) = \sum_{j=\pm 1}  \Delta_A (j,k) = \frac{1}{2} \left ( \sigma_0 + k  \gamma_Y \vec{S}_Y \cdot \vec{\sigma}  \right ) ,
\end{equation}
which can be considered as noisy versions of the operators $X= \vec{S}_X \cdot \vec{\sigma} $ and  $Y= \vec{S}_Y \cdot \vec{\sigma} $. The noise is expressed by the corresponding $\gamma$ factors which are necessary to the joint measurement that would be otherwise forbidden. 

\bigskip

Since this is a noisy simultaneous measurement we look for the general relation between noisy and noiseless statistics. For dichotomic variables there is a simple relation between the noisy marginals $p_{K^\prime}  (\kappa^\prime)$ and the noiseless ones $p_K (\kappa)$, for $K=X,Y$. Taking into account a natural unbiasedness requirement, regarding that if  $p_K (\kappa)$ is uniform then so is $p^\prime_{K^\prime} (\kappa^\prime)$ \cite{YLLO10}, and that $\kappa,\kappa^\prime = \pm 1$, this relation must be of the form
\begin{equation}
\label{nn1}
p^\prime_{K^\prime} (\kappa^\prime |\rho) = \sum_{\kappa = \pm 1}  p_K (\kappa^\prime|\kappa)  p_K (\kappa |\rho) , 
\end{equation}
where the conditional probabilities $p_K (\kappa^\prime|\kappa)$ are
\begin{equation}
\label{nn2}
p_K (\kappa^\prime|\kappa) = \frac{1}{2}  \left  ( 1 +  \gamma_K \kappa \kappa^\prime  \right ) .
\end{equation}
A relevant feature of the schemes we are considering is that the noisy can be easily removed, this is that relations  (\ref{nn1}), (\ref{nn2}) can be inverted to obtain the exact statistics $p_K (\kappa|\rho )$ in terms of the noisy ones $p^\prime_{K^\prime} (\kappa^\prime |\rho)$. That is that for each $K$ there are state-independent functions $\tilde{p}_K (\kappa | \kappa^\prime)$ such that
\begin{equation}
\label{pp1}
p_K (\kappa|\rho ) = \sum_{\kappa^\prime = \pm 1} \tilde{p}_K(\kappa |\kappa^\prime) p^\prime_{K^\prime} (\kappa^\prime |\rho) .
\end{equation}
This inverted distribution must satisfy
\begin{equation}
\sum_{\kappa^\prime} \tilde{p}_K(\kappa_1 |\kappa^\prime) p_K (\kappa^\prime|\kappa_2 ) = \delta_{\kappa_1,\kappa_2} = \frac{1}{2} \left ( 1 + \kappa_1 \kappa_2 \right ) ,
\end{equation}
where the last equality holds for dichotomic variables $\kappa = \pm 1$. After Eq. (\ref{nn2}) we readily get
\begin{equation}
\label{pp2}
p_K (\kappa | \kappa^\prime) = \frac{1}{2} \left ( 1 + \frac{\kappa \kappa^\prime}{ \gamma_K}  \right ) .
\end{equation}

\bigskip
Finally let us note that besides the noisy measurement of observables $X= \vec{S}_X \cdot \vec{\sigma} $ and  $Y= \vec{S}_Y \cdot \vec{\sigma} $ we have also the measurement of $Z=\vec{S}_{XY} \cdot \vec{\sigma}$ contained in the dichotomic variable $xy$. 

\subsection{Marginals, conditional probabilities, and Bell tests}

Some issues of statistics and Bell tests in this probability space have been already addressed in some recent works \cite{MAL20,AL25a,AL25b}, so we refer the interested reader to them. In particular we showed that the violation of Bell inequalities coincides with negative values for the joint distribution for all the measured observables obtained after data inversion removing the noise \cite{MAL20}. We have examined also the satisfaction of Bell criteria for single realizations of the measurement, finding that every outcome violates Bell bounds. This agrees with the idea that to reveal nonclassical effects a necessary condition is that the measuring scheme itself must be nonclassical \cite{AL25a}. Finally we carried out a fully statistical analysis of the results of a Bell test beyond mean values, deriving the probability that a given state violates the bounds \cite{AL25b}.

\section{Conclusions}

In this work we have shown that one and the same experimental arrangement provides a realization of the two very different probability spaces introduced in Refs. \cite{SGH21,SH22}. The possibility of addressing one or the other depends just on the proper labeling of the outcomes, so the two are implemented on the same realization of an experiment providing just one set of raw data.

Regarding probability space 1 we have shown that the scheme provides the random exact measurement of two observables per subsystem. We have suitably analyzed de marginal and conditional statistics showing the possibility of deriving new Bell tests.

\bigskip


\end{document}